\documentclass[12pt,thmsa]{article}
\usepackage{sw20lart}

\input tcilatex
\QQQ{Language}{
British English
}

\begin{document}

\title{Cosmology and the Origin of Life\thanks{%
Invited Talk delivered at the Varenna Conference, \textit{The Origin of
Intelligent Life in the Universe}, 30th September 1998, Varenna.}}
\author{John D. Barrow \\
Astronomy Centre\\
University of Sussex\\
Brighton BN1 9QJ\\
U.K.}
\maketitle
\date{}

\begin{abstract}
We discuss some of the cosmological constraints on the evolution and
persistence of life in the Universe and in hypothetical universes other than
our own. We highlight the role played by the age and size of the universe,
and discuss the interplay between the main-sequence stellar lifetime and the
biological evolution time scale. The consequences of different versions of
the inflationary universe scenario are described in the light of limits on
the possible variation in the values of the constants of Nature.
\end{abstract}

\section{Cosmology, Stars and Life}

Prior to the discovery of the expansion of the Universe there was little
that cosmology could contribute to the question of extraterrestrial life
aside from probabilities and prejudices. After our discovery of the
expansion and evolution of the Universe the situation changed significantly.
The entire cosmic environment was recognised as undergoing steady change.
The history of the Universe took on the complexion of an unfolding drama in
many acts, with the formations of first atoms and molecules, then galaxies
and stars, and most recently, planets and life. The most important and
simplest feature of the overall change in the Universe that the expansion
produces is the rate at which it occurs. This is linked to the age of the
expanding universe and that of its constituents.

In the 1930s, the distinguished biologist JBS Haldane took an interest in
Milne's proposal \cite{milne} that there might exist two different
timescales governing the rates of change of physical processes in the
Universe: one, $t$, for 'atomic' changes and another, $\tau $, for
'gravitational changes' where $\tau =\ln (t/t_0)$ with $t_0$ constant.
Haldane explored how changing from one timescale to the other could alter
ones picture of when conditions in the Universe would become suitable for
the evolution of biochemical life \cite{hald}, \cite{BT}. In particular, he
argued that it would be possible for radioactive decays to occur with a
decay rate that was constant on the $t$ timescale but which grew in
proportion to $t$ when evaluated on the $\tau $ scale. The biochemical
processes associated with energy derived from the breakdown of adenosine
triphosphoric acid would yield energies which, while constant on the $t$
scale, would grow as $t^2$ on the $\tau $ scale. Thus there would be an
epoch of cosmic history on the $\tau $ scale before which life was
impossible but after which it would become increasingly likely. Milne's
theory subsequently fell into abeyance although the interest in gravitation
theories with a varying Newtonian 'constant' of gravitation led to detailed
scrutiny of the paleontological and biological consequences of such
hypothetical changes for the past history of the Earth \cite{BT}.
Ultimately, this led to the formulation of the collection of ideas now known
as the Anthropic Principles, \cite{cart1}, \cite{jb2}.

Another interface between the problem of the origin of life and cosmology
has been the perennial problem of dealing with finite probabilities in
situations where an infinite number of potential trials seem to be
available. For example, in a universe that is infinite in spatial volume (as
would be expected for the case for an expanding open universe with
non-compact topology), any event that has a finite probability of occurring
should occur not just once but infinitely often with probability one if the
spatial structure of the Universe is exhaustively random \cite{ellis}. In
particular, in an infinite universe we conclude that there should exist an
infinite number of sites where life has progressed to our stage of
development. In the case of the steady-state universe, it is possible to
apply this type of argument to the history of the universe as well as its
geography because the universe is assumed to be infinitely old. Every
past-directed world line should encounter a living civilisation.
Accordingly, it has been argued that the steady state universe makes the
awkward prediction that the universe should now be teeming with life along
every line of sight \cite{BT}.

The key ingredient that modern cosmology introduces into considerations of
biology is that of \textit{time}. The observable universe is expanding and
not in a steady state. The density and temperature are steadily falling as
the expansion proceeds. This means that the average ambient conditions in
the universe are linked to its age. Roughly, in all expanding universes,
dimensional analysis tells us that the density of matter, $\rho $, is
related to the age $t$ measured in comoving proper time and Newton's
gravitation constant, $G$, by means of a relation of the form

\begin{equation}
\rho \approx \frac 1{Gt^2}  \label{rho}
\end{equation}

The expanding universe creates an interval of cosmic history during which
biochemical observers, like ourselves, can expect to be examining the
Universe. Chemical complexity requires basic atomic building blocks which
are heavier than the elements of hydrogen and helium which emerge from the
hot early stages of the universe. Heavier elements, like carbon, nitrogen,
and oxygen, are made in the stars, as a result of nuclear reactions that
take billions of years to complete. Then, they are dispersed through space
by supernovae after which they find their way into grains, planets, and
ultimately, into people. This process takes billions of years to complete
and allows the expansion to produce a universe that is billions of light
years in size. Thus we see why it is inevitable that the universe is seen to
be so large. A universe that is billions of years old and hence billions of
light years in size is a necessary pre-requisite for observers based upon
chemical complexity. Biochemists believe that chemical life of this sort,
and the form based upon carbon in particular, is likely to be the only sort
able to evolve spontaneously. Other forms of living complexity (for example
that being sought by means of silicon physics) almost certainly can exist
but it is being developed with carbon-based life-forms as a catalyst rather
than by spontaneous evolution.

The inevitability of universes that are big and old as habitats for life
also leads us to conclude that they must be rather cold on average because
significant expansion to large size reduces the average temperature
inversely in proportion to the size of the universe. They must also be
sparse, with a low average density of matter and large distances between
different stars and galaxies. This low temperature and density also ensures
that the sky is dark at night (the so called 'Olbers' Paradox' first noted
by Halley, \cite{harr}) because there is too little energy available in
space to provide significant apparent luminosity from all the stars. We
conclude that many aspects of our Universe which, superficially, appear
hostile to the evolution of life are necessary prerequisites for the
existence of any form of biological complexity in the Universe.

Life needs to evolve on a timescale that is intermediate between the typical
time scale that it takes for stars to reach a state a state of stable
hydrogen burning, the so called main-sequence lifetime, and the timescale on
which stars exhaust their nuclear fuel and gravitationally collapse. This
timescale, $t_{*}$, is determined by a combination of fundamental constants
of Nature

\begin{equation}
t_{*}\approx \left( \frac{Gm_N^2}{hc}\right) ^{-1}\times \frac h{m_Nc^2}\
\approx 10^9\text{ }yrs  \label{ms}
\end{equation}
where $m_N$ is the proton mass, $h$ is Planck's constant, and $c$ is the
velocity of light \cite{dic}, \cite{BT}.

In expanding universes of the Big Bang type the reciprocal of the observed
expansion rate of the universe, Hubble's constant $H_0\approx
70Km.s^{-1}Mpc^{-1},$ is closely related to the expansion age of the
universe, $t_0$, by a relation of the form

\begin{equation}
t_0\approx \frac 2{3H_0}  \label{H}
\end{equation}
The fact that the age $t_0$ $\approx 10^{10}yr$ deduced from observations of 
$H_0$ in this way is a little larger than the main sequence lifetime, $t_{*}$%
, is entirely natural in the Big Bang theory that is, we observe a little
later than the time when the Sun forms). However, the now defunct steady
state theory, in which there is no relation between the age of the universe
(which is infinite) and the measured value of $H_0,$ would have had to
regard the closeness in value of $H_0^{-1}$ and $t_{*}$ as a complete
coincidence \cite{rees}.

\section{Biology and Stars: Is there a link?}

Evidently, in our solar system life first evolved quite soon after the
formation of a hospitable terrestrial environment. Suppose the typical time
that it takes for life to evolve is denoted by some timescale $\ t_{bio}$,
then from the evidence presented by the solar system, which is about $%
4.6\times 10^9yrs$ old, it is seems that

\[
t_{*}\approx t_{bio} 
\]
At first sight we might assume that the microscopic biochemical processes
and local environmental conditions that combine to determine the magnitude
of $t_{bio}$ are \textit{independent} of the nuclear astrophysical and
gravitational processes that determine the typical stellar main sequence
lifetime $t_{ms}$. However, this assumption leads to the striking conclusion
that we should expect extraterrestrial forms of life to be exceptionally
rare \cite{cart}, \cite{BT}, \cite{les}. The argument, in its simplest form,
is as follows. If $t_{bio}$ and $t_{*\text{ }}$are independent then the time
that life takes to arise is random with respect to the stellar timescale $%
t_{*}$. Thus it is most likely that either $t_{bio}>>t_{*\text{ }}$or that $%
t_{bio}<<t_{*\text{ }}$. Now if$\ t_{bio}<<t_{*\text{ }}$we must ask why it
is that the first observed inhabited solar system (that is, us) has $%
t_{bio}\approx t_{*\text{ }}.$ This would be extraordinarily unlikely. On
the other hand, if $t_{bio}>>t_{*\text{ }}$then the first observed inhabited
solar system (us) is most likely to have $t_{bio}\approx t_{*\text{ }}$since
systems with $t_{bio}>>t_{*\text{ }}$have yet to evolve. Thus we are a
rarity, one of the first living systems to arrive on the scene. Generally,
we are led to a conclusion, an extremely pessimistic one for the SETI
enterprise, that $t_{bio}>>t_{*\text{ }}$.

In order to escape from this conclusion we have to undermine one of the
assumptions underlying the argument that leads to it. For example, if we
suppose that $\ t_{bio}$ is no independent of $t_{*}$ then things look
different. If $t_{bio}/t_{*}$ is a rising function of $t_{*}$ then it is
actually likely that we will find $t_{bio}\approx t_{*\text{ }}$. Livio \cite
{liv} has given a simple model of how it could be that $t_{bio}\ $and $t_{*%
\text{ }}$are related by a relation of this general form. He takes a very
simple model of the evolution of a life-supporting planetary atmosphere like
the Earth's to have two key phases which lead to its oxygen content:

\textit{Phase1}: Oxygen is released by the photodissociation of water
vapour. On Earth this took $2.4\times 10^9yr$ and led to an atmospheric $O_2$
build up to about $10^{-3}\ $of its present value.

\textit{Phase 2}: Oxygen and ozone levels grow to about $0.1$ of their
present levels. This is sufficient to shield the Earth's surface from lethal
levels of ultra-violet radiation in the 2000-3000 \AA\ band (note that
nucleic acid and protein absorption of ultra-violet radiation peaks in the
2600-2700 \AA\ and 2700-2900 \AA\ bands, respectively). On Earth this phase
took about $1.6\times 10^9yr$.

Now the length of Phase 1 might be expected to be inversely proportional to
the intensity of radiation in the wavelength interval 1000-2000 \AA , where
the key molecular levels for $H_2O$ absorption lie. Studies of stellar
evolution allow us to determine this time interval and provide a rough
numerical estimate of the resulting link between the biological evolution
time (assuming it to be determined closely by the photodissociation time)
and the main sequence stellar lifetime, with \cite{liv}

\[
\frac{t_{bio}}{t_{*}}\approx 0.4\left( \frac{t_{*}}{t_{sun}}\right) ^{1.7}, 
\]
where $t_{sun}$ is the age of the Sun.

This model indicates a possible route to establishing a link between the
biochemical timescales for the evolution of life and the astrophysical
timescales that determine the time required to create an environment
supported by a stable hydrogen burning star. There are obvious weak links in
the argument. It provides on a necessary condition for life to evolve, not a
sufficient one. We know that there are many other events that need to occur
before life can evolve in a planetary system. We could imagine being able to
derive an expression for the probability of planet formation around a star.
This would involve many other factors which would determine the amount of
material available for the formation of solid planets with atmospheres at
distances which permit the presence of liquid water and stable surface
conditions. Unfortunately, we know that there were many 'accidents' of the
planetary formation process in the solar system which have subsequently
played a major role in the existence of long-lived stable conditions on
Earth, \cite{art}. For example, the presence of resonances between the
precession rates of rotating planets and the gravitational perturbations
they feel from all other bodies in their solar system can easily produce
chaotic evolution of the tilt of a planet's rotation axis with respect to
the orbital plane of the planets over times must shorter than the age of the
system \cite{tilt}, \cite{art}. The planet's surface temperature variations,
insolation levels, and sea levels are sensitive to this angle of tilt. It
determines the climatic differences between what we call 'the seasons'. In
the case of the Earth, the modest angle of tilt (approximately 23 degrees)
would have experienced this erratic evolution had it not been for the
presence of the Moon \cite{moon}, \cite{art}. The Moon is large enough for
its gravitational effects to dominate the resonances which occur between the
Earth's precessional rotation and the frequency of external gravitational
perturbations from the other planets. As a result the Earth's tilt wobbles
only by a fraction of a degree around $23^{\circ }$ over hundreds of
thousands of years. Enough perhaps to cause some climatic change, but not
catastrophic for the evolution of life.

This shows how the causal link between stellar lifetimes and biological
evolution times may be rather a minor factor in the chain of fortuitous
circumstances that must occur if habitable planets are to form and sustain
viable conditions for the evolution of life over long periods of time. The
problem remains to determine whether he other decisive astronomical factors
in planet formation are functionally linked to the surface conditions needed
for biochemical processes.

\section{Habitable Universes}

We know that several of the distinctive features of the large scale
structure of the visible universe play a role in meeting the conditions
needed for the evolution of biochemical complexity within it.

The first example is the proximity of the expansion dynamics to the
'critical' state which separates an ever-expanding future from one of
eventual contraction, to better than ten per cent. Universes that expanded
far faster than this would be unable to form galaxies and stars and hence
the building blocks of biochemistry would be absent. The rapid expansion
would prevent islands of material separating out from the global expansion
and becoming bound by their own self-gravitation. By contrast, if the
expansion rate were far below that characterising the critical rate then the
material in the universe would have condensed into dense structures and
black holes long before stars could form \cite{ch}, \cite{jb1}, \cite{BT}, 
\cite{jb}, \cite{rees2}.

The second example is that of the uniformity of the universe. The
non-uniformity level on the largest scales is very small, $\Delta \approx
10^{-5}.$ This is a measure of the average relative fluctuations in the
gravitational potential on all scales. If $\Delta $ were significantly
larger then galaxies would have rapidly degenerated into dense structures
within which planetary orbits would be disrupted by tidal forces and black
holes would form rapidly before life-supporting environments could be
established. If $\Delta $ were significantly smaller then the
non-uniformities in the density would be gravitationally too feeble to
collapse into galaxies and no stars would form. Again, the universe would be
bereft of the biochemical building blocks of life \cite{rees3}.

In recent years the most popular theory of the very early evolution of the
universe has provided a possible explanation as to why the universe expands
so close to the critical life-supporting divide and why the fluctuation
level has the value observed. This theory is called 'inflation'. It proposes
that during a short interval of time when the temperature was very high (say 
$\sim 10^{25}K$), the expansion of the universe \textit{accelerated}. This
requires the material content of the universe to be temporarily dominated by
forms of matter which effectively antigravitate for that period of time \cite
{guth}. This requires their density $\rho $, and pressure, $p$, to satisfy
the inequality \cite{jb}

\begin{equation}
\rho +\frac{3p}{c^2}<0  \label{sec}
\end{equation}
The inflation is envisaged to end because the matter fields responsible
decay into other forms of matter, like radiation, which do not satisfy this
inequality. After this occurs the expansion resumes the state of
decelerating expansion that it possessed before its inflationary episode
began.

If inflation occurs it offers the possibility that the whole of the visible
part of the universe (roughly $15$ billion light years in extent today) has
expanded from a region that was small enough to be causally linked by light
signals at the very high temperatures and early times when inflation
occurred. If inflation does not occur then the visible universe would have
expanded from a region that is far larger than the distance that light can
circumnavigate at these early times and so its smoothness today is a
mystery. If inflation occurs it will transform the irreducible quantum
statistical fluctuations in space into distinctive patterns of fluctuations
in the microwave background radiation which future satellite observations
will be able to detect if they were of an intensity sufficient to have
produced the observed galaxies and clusters by the process of gravitational
instability.

As the inflationary universe scenario has been explored in greater depth it
has been found to possess a number of unexpected properties which, if they
are realised, would considerably increase the complexity of the global
cosmological problem and create new perspectives on the existence of life in
the universe \cite{linde}, \cite{vil}, \cite{jb}.

It is possible for inflation to occur in different ways in different places
in the early universe. The effect is rather like the random expansion of a
foam of bubbles. Some inflate considerably while others hardly inflate at
all. This is termed 'chaotic inflation'. Of course, we have to find
ourselves in one of the regions that underwent sufficient inflation so that
the expansion lasted for longer than $t_{*}$ and stars could produce
biological elements. In such a scenario the global structure of the Universe
is predicted to be highly inhomogeneous. Our observations of the microwave
background temperature structure will only be able to tell us whether the
region which expanded to encompass out visible part of the universe
underwent inflation in its past. An important aspect of this theory is that
for the first time it has provided us wit ha positive reason to expect that
the observable universe is not typical of the structure of the universe
beyond our visible horizon, 15 billion light years away.

It was subsequently been discovered that under fairly general conditions
inflation can be self-reproducing. That is, quantum fluctuations within each
inflating bubble will necessarily create conditions for further inflation of
microscopic regions to occur. This process or 'eternal inflation' appears to
have no end and may not have had a beginning. Thus life will be possible
only in bubbles with properties which allow self-organised complexity to
evolve and persist.

It has been found that there is further scope for random variations in these
chaotic and eternal inflationary scenarios. In the standard picture we have
just sketched, properties like the expansion rate and temperature of each
inflated bubble can vary randomly from region to region. However, it is also
possible for the strengths and number of low-energy forces of Nature to
vary. It is even possible for the number of dimensions of space which have
expanded to large size to be different from region to region. We know that
we cannot produce the known varieties of organised biochemical complexity if
the strengths of forces change by relatively small amounts, or in dimensions
other than three because of the impossibility of creating chemical or
gravitational bound states, \cite{eh, whit, tang, BT, teg}.

The possibility of these random variations arises because inflation is ended
by the decay of some matter field satisfying (\ref{sec}). This corresponds
to the field evolving to a minimum in its self-interaction potential. If
that potential has a single minimum then the characteristic physics that
results from that ground state will be the same everywhere. But if the
potential has many minima (for example like a sine function) then each
minimum will have different low-energy physics and different parts of the
universe can emerge from inflation in different minima and with different
effective laws of interaction for elementary particles. In general, we
expect the symmetry breaking which chooses the minima in different regions
to be independent and random.

\section{Changing Constants}

Considerations like these, together with the light that superstring theories
have shed upon the origins of the constants of Nature, mean that we should
assess how narrowly defined the existing constants of Nature need to be in
order to permit biochemical complexity to exist in the Universe \cite{BT}, 
\cite{carr}. For example, if we were to allow the ratio of the electron and
proton masses ($\beta =m_e/m_N$) and the fine structure constant $\alpha $
to be change their values (assuming no other aspects of physics is changed
by this assumption -- which is clearly going to be false!) then the allowed
variations are very constraining. Increase $\beta $ too much and there can
be no ordered molecular structures because the small value of $\beta $
ensures that electrons occupy well-defined positions in the Coulomb field
created by the protons in the nucleus; if $\beta $ exceeds about $5\times
10^{-3}\alpha ^2$ then there would be no stars; if modern grand unified
gauge theories are correct then $\alpha $ must lie in the narrow range
between about $1/180$ and $1/85$ in order that protons not decay too rapidly
and a fundamental unification of non-gravitational forces can occur. If,
instead, we consider the allowed variations in the strength of the strong
nuclear force, $\alpha _s$, and $\alpha $ then roughly $\alpha _s<0.3\alpha
^{1/2}$ is required for the stability of biologically useful elements like
carbon. If we increase $\alpha _s$ by 4\% there is disaster because the
helium-2 isotope can exist (it just fails to be bound by about $70KeV$ in
practice) and allows very fast direct proton + proton $\rightarrow $
helium-2 fusion. Stars would rapidly exhaust their fuel and collapse to
degenerate states or black holes. In contrast, if $\alpha _s$ were decreased
by about 10\% then the deuterium nucleus would cease to be bound and the
nuclear astrophysical pathways to the build up of biological elements would
be blocked. Again, the conclusion is that there is a rather small region of
parameter space in which the basic building blocks of chemical complexity
can exist.

We should stress that conclusions regarding the fragility of living systems
with respect to variations in the values of the constants of Nature are not
fully rigorous in all cases. The values of the constants are simply assumed
to take different constant values to those that they are observed to take
and the consequences of changing them one at a time are examined. However,
if the different constants are fully linked together, as we might expect for
many of them if a unified Theory of Everything exists, then many of these
independent variations may not be possible. The consequences of a small
change in one constant would have further necessary ramifications for the
allowed values of other constants. One would expect the overall effect to be
more constraining on the allowed variations that are life-supporting. For
examples of such coupled variations in string theories see refs. \cite{marc,
barr, drin}.

These considerations are likely to have a bearing on interpreting any future
quantum cosmological theory. Such a theory, by its quantum nature, will make
probabilistic predictions. It will predict that it is 'most probable' that
we find the universe (or its forces and constants) to take particular
values. This presents an interpretational problem because it is not clear
that we should expect the most probable values to be the ones that we
observe. Since only a narrow range of the allowed values for, say, the fine
structure constant will permit observers to exist in the Universe, we must
find ourselves in the narrow range of possibilities which permit them, no
matter how improbable they may be \cite{misha}, \cite{jb}. This means that
in order to fully test the predictions of future Theories of Everything we
must have a thorough understanding of all the ways in which the possible
existence of observers is constrained by variations in the structure of the
universe, in the values of the constants that define its properties, and in
the number of dimensions it possesses. 

\textbf{Acknowledgements}

I would like to thank Professor Elio Sindoni and Donatella Pifferetti for
their efficient organisation and kind hospitality in Varenna and Paul
Davies, Christian de Duve, Mario Livio, Martin Rees, and Max Tegmark for
helpful discussions on some of the topics discussed here. The author was
supported by a PPARC Senior Fellowship.

\end{document}